\documentclass[american,aps,pra,reprint,nofootinbib]{revtex4-1}
\usepackage[T1]{fontenc}
\usepackage[latin9]{inputenc}
\usepackage{color}
\usepackage{babel}
\usepackage{amsmath}
\usepackage{amsthm}
\usepackage{amssymb}
\usepackage{graphicx}
\usepackage[unicode=true,
 bookmarks=false,
 breaklinks=false,pdfborder={0 0 0},pdfborderstyle={},backref=false,colorlinks=true]
 {hyperref}
\hypersetup{
 allcolors=magenta}

\makeatletter
\theoremstyle{plain}
\newtheorem{thm}{\protect\theoremname}

\usepackage{times}
\usepackage{pxfonts}
\usepackage{braket}

\makeatother

\providecommand{\theoremname}{Theorem}

\begin{document}
\title{Quantum state merging with bound entanglement}
\author{Alexander Streltsov}
\email{a.streltsov@cent.uw.edu.pl}

\affiliation{Centre for Quantum Optical Technologies, Centre of New Technologies,
University of Warsaw, Banacha 2c, 02-097 Warsaw, Poland}
\begin{abstract}
Quantum state merging is one of the most important protocols in quantum
information theory. In this task two parties aim to merge their parts
of a pure tripartite state by making use of additional singlets while
preserving correlations with a third party. We study a variation of
this scenario where the shared state is not necessarily pure, and
the merging parties have free access to local operations, classical
communication, and PPT entangled states. We provide general conditions
for a state to admit perfect merging, and present a family of fully
separable states which cannot be perfectly merged if the merging parties
have no access to additional singlets. We also show that free PPT
entangled states do not give any advantage for merging of pure states,
and the conditional entropy plays the same role as in standard quantum
state merging quantifying the rate of additional singlets needed to
perfectly merge the state.
\end{abstract}
\maketitle

\section{Introduction}

Quantum state merging can be understood as a game involving three
players, which we will call Alice, Bob and Charlie in the following.
Initially, they share a large number of copies of a joint pure state
$\ket{\psi}=\ket{\psi}^{ABC}$, and the aim of Bob and Charlie is
to merge their parts of the state on Charlie's side while preserving
correlations with Alice. For achieving this, Bob and Charlie have
access to additional singlets and a classical communication channel.
Taking into account that singlets are considered as an expensive resource
in quantum information theory, the main question of quantum state
merging can be formulated as follows: How many singlets are required
for perfect asymptotic merging per copy of the state\emph{ $\ket{\psi}$}?
The answer to this question was found in \citep{Horodecki2005,Horodecki2007}:
the minimal number of singlets per copy is given by the conditional
entropy $S(\rho^{BC})-S(\rho^{C})$. 

Noting that the conditional entropy can be positive or negative, it
is surprising that it admits an operational interpretation in both
cases. In particular, if the conditional entropy is positive, Bob
and Charlie will require $S(\rho^{BC})-S(\rho^{C})$ singlets per
copy for perfectly merging the total state $\ket{\psi}$ in the asymptotic
limit, and perfect merging cannot be accomplished if less singlets
are available. On the other hand, if $S(\rho^{BC})-S(\rho^{C})$ is
negative, Bob and Charlie can asymptotically merge the state $\ket{\psi}$
without any additional singlets by only using local operations and
classical communication (LOCC). Moreover, Bob and Charlie can gain
additional singlets at rate $S(\rho^{C})-S(\rho^{BC})$, and store
them for future use \citep{Horodecki2005,Horodecki2007}.

Another important concept in quantum information theory is the framework
of entanglement distillation \citep{Bennett1996a,Bennett1996,Horodecki2009}.
One of the most surprising features in this context is the phenomenon
of bound entanglement: there exist entangled states from which no
singlets can be distilled~\citep{Horodecki1998}. Moreover, it is
known that all states with positive partial transpose (PPT) are nondistillable
\citep{Horodecki1998}, while it is still an open question if there
exist bound entangled states with nonpositive partial transpose (NPT)
\citep{Pankowski2010}. 

In this paper we introduce and study the task of \emph{PPT quantum
state merging} \emph{(PQSM)}. Similar to standard quantum state merging,
PQSM can be considered as a game between three players who share a
joint mixed state $\rho=\rho^{ABC}$. The aim of the game for Bob
and Charlie is to merge their parts of the state $\rho$ on Charlie's
side while preserving correlations with Alice. In contrast to standard
quantum state merging, Bob and Charlie can use unlimited amount of
PPT entangled states, see Fig.~\ref{fig:setup} for illustration.
The situation where Bob and Charlie do not have access to PPT entangled
states is known as LOCC quantum state merging (LQSM), and has been
introduced in \citep{Streltsov2015}.

Before we discuss the concept of PPT quantum state merging and present
our main results, we will introduce PPT assisted LOCC operations in
the following.

\section{PPT assisted LOCC}

\begin{figure}
\includegraphics[width=1\columnwidth]{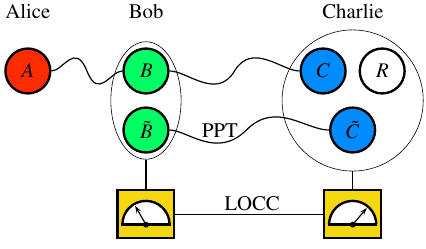}

\caption{\label{fig:setup}PPT quantum state merging (PQSM). Alice, Bob and
Charlie initially share a joint state $\rho=\rho^{ABC}$. Bob and
Charlie aim to merge Bob's part of $\rho$ on Charlie's side, while
preserving correlations with Alice. For this, Bob and Charlie have
access to arbitrary PPT states $\mu_{\mathrm{PPT}}=\mu_{\mathrm{PPT}}^{\tilde{B}\tilde{C}}$,
and can perform local operations on their parts and communicate the
outcomes via a classical channel. The register $R$ in Charlie's hands
serves as storage: in the ideal case, the final state $\sigma_{f}^{ACR}$
is equivalent to $\rho^{ABC}$ up to relabeling $B$ and $R$.}
\end{figure}
For a tripartite state $\rho^{ABC}$ shared between Alice, Bob and
Charlie, a PPT assisted LOCC protocol performed by Charlie and Bob
will be denoted by $\Lambda_{\mathrm{PPT}}$ and has the following
form: 
\begin{equation}
\Lambda_{\mathrm{PPT}}\left(\rho^{ABC}\right)=\mathrm{Tr}_{\tilde{B}\tilde{C}}\left[\Lambda_{\mathrm{LOCC}}\left[\rho^{ABC}\otimes\mu_{\mathrm{PPT}}^{\tilde{B}\tilde{C}}\right]\right].
\end{equation}
Here, $\mu_{\mathrm{PPT}}^{\tilde{B}\tilde{C}}$ is an arbitrary PPT
state shared by Bob and Charlie, and $\Lambda_{\mathrm{LOCC}}$ is
an LOCC protocol between them, see Fig.~\ref{fig:setup}.

We also introduce PPT distillable entanglement $D_{\mathrm{PPT}}$
as the singlet rate which can be asymptotically obtained from a state
via PPT assisted LOCC. This quantity is in full analogy to the standard
distillable entanglement \citep{Bennett1996a} that quantifies the
singlet rate which can be obtained via LOCC only. We will denote the
latter by $D_{\mathrm{LOCC}}$. 

If we further introduce the PPT and LOCC entanglement cost $C_{\mathrm{PPT}}$
and $C_{\mathrm{LOCC}}$ as the entanglement cost for creating a state
via the corresponding set of operations, we immediately obtain the
following inequality: 
\begin{equation}
D_{\mathrm{LOCC}}(\rho)\leq D_{\mathrm{PPT}}(\rho)\leq C_{\mathrm{PPT}}(\rho)\leq C_{\mathrm{LOCC}}(\rho).\label{eq:inclusions}
\end{equation}
This relation follows by noting that PPT assisted LOCC is more general
than LOCC only, and by the fact that the PPT entanglement cost cannot
be below the PPT distillable entanglement. Since for all pure states
$D_{\mathrm{LOCC}}$ and $C_{\mathrm{LOCC}}$ are equal to the von
Neumann entropy of the reduced state~\citep{Bennett1996}, all quantities
in Eq.~(\ref{eq:inclusions}) coincide for pure states. In the following,
we will also use the logarithmic negativity \citep{Zyczkowski1998,Vidal2002}
\begin{equation}
E_{n}(\rho)=\log_{2}||\rho^{T_{A}}||,
\end{equation}
where $T_{A}$ denotes partial transposition, and $||M||=\mathrm{Tr}\sqrt{M^{\dagger}M}$
is the trace norm of $M$. The logarithmic negativity is an upper
bound on $D_{\mathrm{LOCC}}$ \citep{Vidal2002}.

We further note that PPT assisted LOCC is a subclass of general PPT
preserving operations. It is however not clear whether or not these
two classes coincide.

\section{PPT quantum state merging}

We are now in position to introduce the aforementioned task of PPT
quantum state merging (PQSM). In this task, Bob and Charlie aim to
merge their parts of the total state $\rho=\rho^{ABC}$ by using PPT
assisted LOCC operations, see Fig.~\ref{fig:setup} for illustration.
A natural figure of merit for this process is the fidelity of PQSM:
\begin{equation}
{\cal F}_{\mathrm{PPT}}(\rho)=\sup_{\Lambda_{\mathrm{PPT}}}F(\sigma_{f},\sigma_{t})\label{eq:fidelity}
\end{equation}
with fidelity $F(\rho,\sigma)=\mathrm{Tr}(\sqrt{\rho}\sigma\sqrt{\rho})^{1/2}$.
In the above expression, the target state $\sigma_{t}=\sigma_{t}^{ACR}$
is the same as $\rho=\rho^{ABC}$ up to relabeling of the systems
$B$ and $R$, where $R$ is an additional register in Charlie's hands.
The final state $\sigma_{f}=\sigma_{f}^{ACR}$ shared by Alice and
Charlie is given by 
\begin{equation}
\sigma_{f}=\mathrm{Tr}_{B}\left[\Lambda_{\mathrm{PPT}}\left[\rho^{ABC}\otimes\rho^{R}\right]\right],\label{eq:final}
\end{equation}
where $\rho^{R}$ is an arbitrary initial state of Charlie's register
$R$. The supremum in Eq.~(\ref{eq:fidelity}) is taken over all
PPT assisted LOCC operations $\Lambda_{\mathrm{PPT}}$ between Bob's
system $B$ and Charlie's system $CR$, see also Fig.~\ref{fig:setup}
for details. A state $\rho$ admits \emph{perfect single-shot PQSM}
if and only if the corresponding fidelity is equal to one: ${\cal F}_{\mathrm{PPT}}(\rho)=1,$
and ${\cal F}_{\mathrm{PPT}}(\rho)<1$ otherwise. 

In the asymptotic scenario where a large number of copies of the state
$\rho$ is available, the figure of merit is the asymptotic fidelity
of PQSM: 
\begin{equation}
{\cal F}_{\mathrm{PPT}}^{\infty}(\rho)=\lim_{n\rightarrow\infty}{\cal F}_{\mathrm{PPT}}(\rho^{\otimes n}).
\end{equation}
This quantity can be regarded as a natural quantifier for asymptotic
PQSM, since a state $\rho$ admits \emph{perfect asymptotic PQSM}
if and only if ${\cal F}_{\mathrm{PPT}}^{\infty}(\rho)=1$.

\section{Perfect asymptotic PQSM}

In the following we will focus on those states $\rho=\rho^{ABC}$
which admit perfect asymptotic PQSM:
\begin{equation}
{\cal F}_{\mathrm{PPT}}^{\infty}(\rho)=1.\label{eq:perfect-1}
\end{equation}
In particular, perfect asymptotic PQSM is always possible if the state
$\rho$ has nonpositive conditional entropy: 
\begin{equation}
S(\rho^{BC})-S(\rho^{C})\leq0.\label{eq:sufficient}
\end{equation}
This follows from the fact that in this situation Bob and Charlie
can achieve perfect asymptotic merging for the purification of $\rho$
just by using local operations and classical communication \citep{Horodecki2005,Horodecki2007,Streltsov2015}.
Moreover, Eq.~(\ref{eq:sufficient}) also implies that states satisfying
Eq.~(\ref{eq:perfect-1}) have nonzero measure in the set of all
states, since this is evidently true for states satisfying Eq.~(\ref{eq:sufficient}). 

At this point, we also note that perfect asymptotic PQSM is only possible
if the state $\rho$ satisfies the following condition:
\begin{equation}
D_{\mathrm{PPT}}^{A:BC}(\rho)\leq D_{\mathrm{PPT}}^{AB:C}(\rho),\label{eq:En-1}
\end{equation}
where $X:Y$ denotes a bipartition between two (possibly multipartite)
subsystems $X$ and $Y$. To see this, consider the overall state
$\rho^{ABC}\otimes\rho^{R}$, where $R$ is a register in Charlie's
hands. If this state allows for perfect asymptotic PQSM, there exists
a PPT assisted LOCC protocol $\Lambda_{\mathrm{PPT}}$ between Bob
and Charlie such that 
\begin{equation}
\rho^{ABC}\otimes\rho^{R}\overset{\Lambda_{\mathrm{PPT}}}{\longrightarrow}\rho^{ACR}\otimes\ket{0}\!\bra{0}^{B}.
\end{equation}
Consider now the PPT distillable entanglement in the bipartition $AB:CR$.
By its very definition, PPT distillable entanglement cannot grow under
PPT assisted LOCC operations, and we obtain 
\begin{equation}
D_{\mathrm{PPT}}^{AB:CR}(\rho^{ACR}\otimes\ket{0}\!\bra{0}^{B})\leq D_{\mathrm{PPT}}^{AB:CR}(\rho^{ABC}\otimes\rho^{R}).
\end{equation}
Noting that the states $\rho^{ABC}$ and $\rho^{ACR}$ differ only
by relabeling $B$ and $R$ completes the proof of Eq.~(\ref{eq:En-1}).

For states which satisfy Eq.~(\ref{eq:En-1}) but violate Eq.~(\ref{eq:sufficient})
no conclusive statement can be made in general. One important subclass
of such states are fully separable states, and it is easy to provide
examples for such states which violate Eq.~(\ref{eq:sufficient}),
but still can be merged via LOCC even on the single-copy level. In
the following we will show that the investigation of such states can
be simplified significantly. This will also lead us to a new class
of fully separable states which cannot be merged via asymptotic PQSM.

\section{Single-shot versus asymptotic PQSM}

In the following, we consider the situation where the total state
$\rho=\rho^{ABC}$ has positive partial transpose with respect to
the bipartition $AB$:$C$. The set of these states includes the aforementioned
set of fully separable states. The following theorem shows that for
all such states the single-copy fidelity is never smaller than for
any number of copies.
\begin{thm}
\label{thm:1}Given a tripartite state $\rho=\rho^{ABC}$ which is
PPT with respect to $AB$:$C$, the following inequality holds for
any $n\geq1$: 
\begin{equation}
{\cal F}_{\mathrm{PPT}}(\rho)\geq{\cal F}_{\mathrm{PPT}}(\rho^{\otimes n}).
\end{equation}
\end{thm}
\noindent This also implies that in this case the single-shot fidelity
cannot be smaller than the asymptotic fidelity: ${\cal F}_{\mathrm{PPT}}(\rho)\geq{\cal F}_{\mathrm{PPT}}^{\infty}(\rho)$.
We refer to the Appendix for the proof.

Crucially, this result also means that perfect single-shot PQSM is
fully equivalent to perfect asymptotic PQSM for all such states: 
\begin{equation}
{\cal F}_{\mathrm{PPT}}(\rho)=1\Leftrightarrow{\cal F}_{\mathrm{PPT}}^{\infty}(\rho)=1.\label{eq:perfect}
\end{equation}
The importance of this result lies in the fact that it remarkably
simplifies the analysis, if one is interested in the question whether
a state $\rho$ admits perfect asymptotic PQSM or not. For all such
states we only need to study the single-copy situation: if perfect
PQSM is not possible in the single-copy case, it is also not possible
asymptotically.

As an application of Theorem \ref{thm:1}, we will now present a general
family of fully separable states which does not admit perfect asymptotic
PQSM. These states are given by 
\begin{align}
\rho_{\mathrm{sep}}^{ABC} & =\sum_{i=0}^{14}p_{i}\ket{i}\bra{i}^{A}\otimes\sigma_{i}^{BC},\label{eq:nomerging}
\end{align}
where all probabilities $p_{i}$ are nonzero, and the two-qubit states
$\sigma_{i}^{BC}$ are all separable and chosen such that their generalized
Bloch vectors are all linearly independent. For the proof that such
states exist and that they indeed do not allow for perfect asymptotic
PQSM we refer to the Appendix.

\section{States with vanishing asymptotic fidelity}

Taking into account the results discussed so far, it is natural to
ask whether the asymptotic fidelity ${\cal F}_{\mathrm{PPT}}^{\infty}$
can attain only one of two values, namely 0 or 1. We can neither prove
nor disprove this at the moment. Nevertheless, we will provide strong
evidence for this in the following, showing that a significant amount
of quantum states has vanishing asymptotic fidelity: 
\begin{equation}
{\cal F}_{\mathrm{PPT}}^{\infty}(\rho)=0.\label{eq:vanishing}
\end{equation}
This happens for all states which are distillable between $A$ and
$BC$, and at the same time have positive partial transpose in the
bipartition $AB$:$C$. These two conditions are summarized in the
following inequality:
\begin{equation}
D_{\mathrm{LOCC}}^{A:BC}(\rho)>E_{n}^{AB:C}(\rho)=0.\label{eq:vanishing-1}
\end{equation}
The proof of this statement can be found in the Appendix.

At this point it is also interesting to note that the asymptotic fidelity
${\cal F}_{\mathrm{PPT}}^{\infty}$ is not a continuous function of
the state. This discontinuity is present even for pure states, and
can be demonstrated on the following example: 
\begin{equation}
\rho=\ket{\psi}\bra{\psi}^{AB}\otimes\ket{0}\bra{0}^{C}.\label{eq:example}
\end{equation}
Note that this state admits perfect PQSM whenever $\ket{\psi}$ is
a product state, i.e., $\ket{\psi}=\ket{\alpha}^{A}\otimes\ket{\beta}^{B}$.
In this case, perfect merging can be accomplished without any communication
if Charlie prepares his register $R$ in the state $\ket{\beta}^{R}$.
Note however that the asymptotic fidelity ${\cal F}_{\mathrm{PPT}}^{\infty}$
vanishes for any entangled state $\ket{\psi}$, as follows directly
from the above discussion. 

As is further shown in the Appendix, the set of states having vanishing
asymptotic fidelity has nonzero measure in the set of all states.
Combining these results with our previous findings, namely that states
satisfying ${\cal F}_{\mathrm{PPT}}^{\infty}(\rho)=1$ also have nonzero
measure in the set of all states, this means that both of these sets
have finite size. We hope that this result can serve as a starting
point to prove that ${\cal F}_{\mathrm{PPT}}^{\infty}$ can take as
values only $0$ or $1$.

\section{Absence of bound entanglement}

The results presented in this work can also be applied to the scenario
where Bob and Charlie do not have access to PPT entangled states.
This task is known as LOCC quantum state merging (LQSM), and has been
presented in \citep{Streltsov2015}. The figure of merit in this case
will be denoted by ${\cal F}_{\mathrm{LOCC}}$. 

Note that the quantities ${\cal F}_{\mathrm{LOCC}}$ and ${\cal F}_{\mathrm{PPT}}$
obey the following relation: 
\begin{equation}
{\cal F}_{\mathrm{PPT}}(\rho)\geq{\cal F}_{\mathrm{LOCC}}(\rho)\geq2^{\frac{1}{2}\left[{\cal I}(\rho)-I^{A:BC}(\rho)\right]}.\label{eq:LOCCvsPPT}
\end{equation}
Here, $I^{A:BC}$ is the mutual information between $A$ and $BC$,
and ${\cal I}$ is the concentrated information introduced in~\citep{Streltsov2015}.
The concentrated information quantifies the maximal amount of mutual
information between Alice and Charlie obtainable via LOCC operations
performed by Charlie and Bob, and can be considered as a figure of
merit for LQSM on its own right. The first inequality in~(\ref{eq:LOCCvsPPT})
follows from the fact that PPT assisted LOCC operations are more general
than LOCC operations alone. The second inequality in (\ref{eq:LOCCvsPPT})
crucially relies on results from \citep{Fawzi2014,Sutter2015,Wilde2015},
and the proof can be found in \citep{Streltsov2015}.

The second inequality in (\ref{eq:LOCCvsPPT}) further implies that
${\cal F}_{\mathrm{LOCC}}$ and ${\cal F}_{\mathrm{PPT}}$ are nonzero
for any finite-dimensional state $\rho$. This follows directly by
noting that the concentrated information ${\cal I}$ is nonnegative,
and that the mutual information $I^{A:BC}$ is finite. The first inequality
in (\ref{eq:LOCCvsPPT}) implies that all states with vanishing asymptotic
PQSM fidelity also have zero asymptotic LQSM fidelity: ${\cal F}_{\mathrm{PPT}}^{\infty}(\rho)=0$
implies ${\cal F}_{\mathrm{LOCC}}^{\infty}(\rho)=0$. This means that
all states $\rho$ which fulfill Eq.~(\ref{eq:vanishing-1}) also
have vanishing asymptotic LQSM fidelity: ${\cal F}_{\mathrm{LOCC}}^{\infty}(\rho)=0$. 

This result can be slightly generalized by using the same arguments
as in the proof of Eq.~(\ref{eq:vanishing}). In particular, all
states $\rho$ which are distillable between $A$ and $BC$ but nondistillable
with respect to $AB$:$C$ have vanishing asymptotic fidelity for
LQSM, i.e., 
\begin{equation}
D_{\mathrm{LOCC}}^{A:BC}(\rho)>D_{\mathrm{LOCC}}^{AB:C}(\rho)=0\label{eq:LOCC}
\end{equation}
implies ${\cal F}_{\mathrm{LOCC}}^{\infty}(\rho)=0$. For proving
this, we can use the same proof as for Eq.~(\ref{eq:vanishing}),
by noting that the final state shared by Alice and Charlie will never
be distillable if the initial state $\rho$ satisfies Eq.~(\ref{eq:LOCC}),
and if Bob and Charlie use LOCC operations only. 

At this point we also note that Eq. (\ref{eq:LOCC}) does not guarantee
vanishing PQSM fidelity. In particular, if there exist NPT bound entangled
states -- and it is strongly believed that this is indeed the case
\citep{Pankowski2010} -- Bob and Charlie could use PPT entangled
states to perfectly merge a state of the form 
\begin{equation}
\rho=\ket{\phi^{+}}\bra{\phi^{+}}^{AB_{1}}\otimes\rho_{\mathrm{NPT}}^{B_{2}C},
\end{equation}
where the particles $B_{1}$ and $B_{2}$ are in Bob's hands, $\ket{\phi^{+}}=(\ket{00}+\ket{11})/\sqrt{2}$
is a maximally entangled two-qubit state, and $\rho_{\mathrm{NPT}}$
is an NPT bound entangled state with the property that $D_{\mathrm{LOCC}}(\rho_{\mathrm{NPT}}\otimes\mu_{\mathrm{PPT}})>1$
for some PPT entangled state $\mu_{\mathrm{PPT}}$. Note that states
$\rho_{\mathrm{NPT}}$ and $\mu_{\mathrm{PPT}}$ with the aforementioned
properties exist if there are NPT bound entangled states \citep{Vollbrecht2002}.
Bob and Charlie can then use the state $\mu_{\mathrm{PPT}}$ to distill
the state $\rho_{\mathrm{NPT}}\otimes\mu_{\mathrm{PPT}}$, and by
applying Schumacher compression \citep{Schumacher1995} to achieve
${\cal F}_{\mathrm{PPT}}^{\infty}(\rho)=1$.

We also note that all states $\rho$ which fulfill the condition~(\ref{eq:sufficient})
admit perfect asymptotic LQSM \citep{Streltsov2015}, which also implies
that states with ${\cal F}_{\mathrm{LOCC}}^{\infty}(\rho)=1$ have
nonzero measure in the set of all states. Moreover, a state $\rho$
admits perfect asymptotic LQSM only if it satisfies the following
condition: 
\begin{equation}
D_{\mathrm{LOCC}}^{A:BC}(\rho)\leq D_{\mathrm{LOCC}}^{AB:C}(\rho).\label{eq:Ed}
\end{equation}
Similar to the condition (\ref{eq:En-1}) for perfect asymptotic PQSM,
Eq. (\ref{eq:Ed}) follows from the fact that distillable entanglement
cannot increase under LOCC operations. 

\section{Comparison to standard quantum state merging}

In the setting discussed so far we assumed that Bob and Charlie have
free access to PPT entangled states together with local operations
and classical communication. To compare our results to standard quantum
state merging~\citep{Horodecki2005,Horodecki2007}, we will now also
allow Bob and Charlie to share singlets. The main question of this
section can be formulated as follows: \emph{can shared PPT states
reduce the singlet rate required for merging?} As we will see in the
following, the answer to this question is negative: also in the presence
of PPT states the minimal singlet rate needed to achieve perfect merging
of a tripartite pure state $\ket{\psi}^{ABC}$ corresponds to the
conditional entropy $S(\rho^{BC})-S(\rho^{C})$.

If Bob and Charlie have access to additional entangled states $\ket{D_{i}}^{B'C'}$
with initial distillable entanglement $D_{i}$, perfect PQSM of the
state $\ket{\psi}=\ket{\psi}^{ABC}$ can be seen as the following
asymptotic transformation:
\begin{equation}
\ket{\psi}^{ABC}\otimes\ket{D_{i}}^{B'C'}\otimes\ket{0}^{R}\overset{\Lambda_{\mathrm{PPT}}}{\longrightarrow}\ket{\psi}^{ACR}\otimes\ket{D_{f}}^{B'C'}\otimes\ket{0}^{B}.\label{eq:purestatemerging}
\end{equation}
Here, $R$ is a register in Charlie's possession, and the state $\ket{D_{f}}^{B'C'}$
has final distillable entanglement $D_{f}$. This condition means
that by using additional singlets at rate $D_{i}$, Bob and Charlie
can perfectly merge the state $\ket{\psi}$ in the asymptotic limit
via PPT assisted LOCC, and will at the same time gain singlets at
rate $D_{f}$. The entanglement cost of the process is then given
by $D_{i}-D_{f}$.

We will now show that the conditional entropy of the reduced state
$\rho^{BC}$ is equal to the minimal entanglement cost of the above
process. For this, we note that perfect merging is always possible
at cost $D_{i}-D_{f}=S(\rho^{BC})-S(\rho^{C})$, since there exists
an LOCC protocol accomplishing this task at this cost \citep{Horodecki2005,Horodecki2007}.
In the following, we will see that PPT assisted LOCC cannot lead to
lower cost, i.e., 
\begin{align}
D_{i}-D_{f} & \geq S(\rho^{BC})-S(\rho^{C})\label{eq:bound}
\end{align}
is true for any PPT assisted LOCC protocol achieving perfect merging
as in Eq.~(\ref{eq:purestatemerging}). For proving this, we will
introduce the initial state $\ket{\Psi_{i}}$ and the final state
$\ket{\Psi_{f}}$. They correspond to the total state on the left-hand
side and the right-hand side of Eq.~(\ref{eq:purestatemerging}),
respectively. Using the fact that for pure states the PPT distillable
entanglement $D_{\mathrm{PPT}}$ is equal to the von Neumann entropy
of the reduced state (see also Eq.~(\ref{eq:inclusions}) and discussion
there), it is straightforward to verify the following equality: 
\begin{align}
D_{i}-D_{f} & =S(\rho^{BC})-S(\rho^{C})+D_{\mathrm{PPT}}(\ket{\Psi_{i}})-D_{\mathrm{PPT}}(\ket{\Psi_{f}}),
\end{align}
where the PPT distillable entanglement $D_{\mathrm{PPT}}$ is considered
with respect to the bipartition $ABB'$:$CC'R$. The desired inequality
(\ref{eq:bound}) follows by noting that $D_{\mathrm{PPT}}$ cannot
increase under PPT assisted LOCC, and thus $D_{\mathrm{PPT}}(\ket{\Psi_{i}})\geq D_{\mathrm{PPT}}(\ket{\Psi_{f}})$.\\

\section{Conclusions}

In this paper we introduced and studied the task of PPT quantum state
merging (PQSM), where two parties -- Bob and Charlie -- aim to merge
their shares of a tripartite mixed state by using PPT entanglement
and classical communication, while preserving correlations with Alice. 

We considered the fidelity of this process, both in the single-copy
and the asymptotic scenario, and showed that fully separable states
can be perfectly merged asymptotically if and only if they can be
perfectly merged on the single-copy level. We used this result to
present a family of fully separable states which do not admit perfect
asymptotic PQSM. We also identified very general conditions for a
state to have vanishing fidelity of PQSM in the asymptotic limit.
We showed that these conditions apply to a significant amount of quantum
states having nonzero measure in the set of all states, thus proving
that a large number of quantum states cannot be merged asymptotically
with any nonzero precision. With respect to standard quantum state
merging, our results imply that using additional PPT states does not
change the entanglement cost of the process: the minimal singlet rate
needed for perfectly merging a pure state in the asymptotic limit
corresponds to the conditional entropy also in this extended setup. 

We further note that the protocol considered here cannot be extended
to the scenario where Bob and Charlie have access to arbitrary bound
entangled states. In particular, if there exist NPT bound entangled
states, the results presented in \citep{Vollbrecht2002} immediately
imply that Bob and Charlie also have access to an unlimited amount
of singlets, and thus all states can be perfectly merged. On the other
hand, if NPT bound entangled states do not exist, the scenario described
here already represents the most general situation.

We expect that the tools presented here will find applications for
other quantum communication protocols such as quantum state redistribution~\citep{DevetakPhysRevLett.100.230501},
also taking into account possible local constraints~\citep{StreltsovPhysRevLett.116.240405,Anshu1804.04915}.
However, these questions are beyond the scope of this work.
\begin{acknowledgments}
I acknowledge discussion with Andreas Winter. This work was supported
by the Alexander von Humboldt-Foundation and the ``Quantum Coherence
and Entanglement for Quantum Technology'' project, carried out within
the First Team programme of the Foundation for Polish Science co-financed
by the European Union under the European Regional Development Fund.
\end{acknowledgments}

\appendix

\section{Proof of Theorem 1}

In the following we will prove that any state $\rho=\rho^{ABC}$ which
is PPT with respect to the bipartition $AB:C$ satisfies the inequality
\begin{equation}
{\cal F}_{\mathrm{PPT}}(\rho)\geq{\cal F}_{\mathrm{PPT}}(\rho^{\otimes n})\label{eq:separable}
\end{equation}
 for any number of copies $n\geq1$. We will prove this inequality
for $n=2$, and for larger $n$ the proof follows similar lines of
reasoning. 

For $n=2$ we will denote the total initial state by 
\begin{equation}
\rho\otimes\rho=\rho^{A_{1}B_{1}C_{1}}\otimes\rho^{A_{2}B_{2}C_{2}},
\end{equation}
and the final state $\sigma_{f}=\sigma_{f}^{A_{1}A_{2}C_{1}C_{2}R_{1}R_{2}}$
is then given by 
\begin{equation}
\sigma_{f}=\mathrm{Tr}_{B_{1}B_{2}\tilde{B}\tilde{C}}\left[\Lambda\left[\rho^{A_{1}B_{1}C_{1}}\otimes\rho^{A_{2}B_{2}C_{2}}\otimes\mu_{\mathrm{PPT}}^{\tilde{B}\tilde{C}}\otimes\rho^{R_{1}R_{2}}\right]\right],\label{eq:rhof}
\end{equation}
where $\mu_{\mathrm{PPT}}^{\tilde{B}\tilde{C}}$ is a PPT state, $\Lambda$
is an LOCC operation between Bob's total system $B_{1}B_{2}\tilde{B}$
and Charlie's total system $C_{1}C_{2}\tilde{C}R_{1}R_{2}$, and $\rho^{R_{1}R_{2}}$
is an arbitrary initial state of Charlie's register. 

We will now prove Eq.~(\ref{eq:separable}) by contradiction, assuming
that it is violated for some state $\rho$ which is PPT with respect
to $AB:C$. In this case there must exist a PPT state $\mu_{\mathrm{PPT}}$
and an LOCC protocol $\Lambda$ such that
\begin{equation}
F(\sigma_{f},\sigma_{t}^{A_{1}C_{1}R_{1}}\otimes\sigma_{t}^{A_{2}C_{2}R_{2}})>{\cal F}_{\mathrm{PPT}}(\rho),\label{eq:two copies}
\end{equation}
where the final state $\sigma_{f}$ was given in Eq.~(\ref{eq:rhof}).
The target state $\sigma_{t}^{A_{1}C_{1}R_{1}}\otimes\sigma_{t}^{A_{2}C_{2}R_{2}}$
is the same as $\rho^{A_{1}B_{1}C_{1}}\otimes\rho^{A_{2}B_{2}C_{2}}$
up to relabeling the parties $B_{1}$ and $R_{1}$, and $B_{2}$ and
$R_{2}$.

\begin{figure}
\includegraphics[width=1\columnwidth]{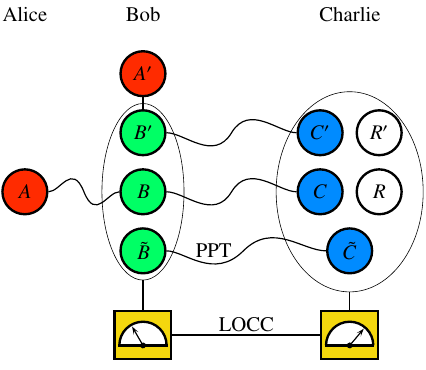}

\caption{\label{fig:2} Proof of Eq.~(\ref{eq:separable}) for $n=2$. A violation
of Eq.~(\ref{eq:separable}) could be used to build a protocol acting
on one copy of the state $\rho$, and reaching a higher single-copy
fidelity than ${\cal F}_{\mathrm{PPT}}(\rho)$.}
\end{figure}
We will now show that Bob and Charlie can ``simulate'' such a two-copy
protocol with just one copy of the state $\rho$, thus achieving a
single-copy fidelity strictly above ${\cal F}_{\mathrm{PPT}}$, which
will be the desired contradiction. The basic idea of the proof is
illustrated in Fig.~\ref{fig:2}. We assume that Alice, Bob and Charlie
start with only one copy of the state $\rho=\rho^{ABC}$, and that
the state is PPT between $AB$ and $C$. Since Bob and Charlie can
prepare arbitrary PPT states, they can additionally prepare the state
$\rho^{A'B'C'}$, which is equivalent to $\rho^{ABC}$ up to the fact
that $A'$ and $B'$ are both in Bob's possession, see Fig. \ref{fig:2}.

In the next step, Bob and Charly prepare a PPT state $\mu_{\mathrm{PPT}}$
and run the same LOCC protocol $\Lambda$ which was leading to Eq.~(\ref{eq:two copies}).
By following this strategy, they will end up with a final state $\sigma_{f}^{AA'CC'RR'}$
having the property that 
\begin{equation}
F(\sigma_{f}^{AA'CC'RR'},\sigma_{t}^{ACR}\otimes\sigma_{t}^{A'C'R'})>{\cal F}_{\mathrm{PPT}}(\rho).
\end{equation}
Recalling that fidelity does not decrease under discarding subsystems,
it follows that 
\begin{equation}
F(\sigma_{f}^{ACR},\sigma_{t}^{ACR})>{\cal F}_{\mathrm{PPT}}(\rho),
\end{equation}
which is the desired contradiction.

The proof for arbitrary $n\geq2$ follows by applying the same arguments.
Moreover, using the same ideas it is possible to show that the fidelity
of LQSM ${\cal F}$ satisfies the inequality 
\begin{equation}
{\cal F}(\rho)\geq{\cal F}(\rho^{\otimes n})
\end{equation}
for any $n\geq2$ and any state $\rho$ which is separable between
$AB$ and $C$.

\section{Fully separable states not admitting perfect asymptotic PQSM}

Here we will present a family of fully separable tripartite states
$\rho_{\mathrm{sep}}^{ABC}$ that cannot be merged via PPT assisted
LOCC even in the asymptotic scenario. The desired family of states
is given by
\begin{align}
\rho_{\mathrm{sep}}^{ABC} & =\sum_{i=0}^{14}p_{i}\ket{i}\bra{i}^{A}\otimes\sigma_{i}^{BC}.\label{eq:nomerging-1}
\end{align}
Here, all states $\sigma_{i}^{BC}$ are separable two-qubit states
and the particle $A$ has dimension $15$ (the reason for this will
become clear below). The probabilities $p_{i}$ are strictly positive
for all $0\leq i\leq14$. 

Note that any general $d$-dimensional Hilbert space has an associated
Bloch vector space of dimension $d^{2}-1$~\citep{Bertlmann2008}.
In the case considered here, the particles $B$ and $C$ are qubits.
Thus, the Bloch vector space associated with the Hilbert space of
$BC$ has dimension $15$. Moreover, note that there exist $15$ separable
two-qubit states $\sigma_{i}^{BC}$ with the property that all their
Bloch vectors are linearly independent. This follows from the fact
that the set of separable states has finite size within the set of
all states~\citep{Gurvits2003}.

As we will see in the following, the state in Eq.~(\ref{eq:nomerging-1})
cannot be merged via PPT assisted LOCC whenever the generalized Bloch
vectors of the states $\sigma_{i}^{BC}$ are linearly independent
for all $0\leq i\leq14$. Due to Theorem~\ref{thm:1} of the main
text it is enough to focus on the single-shot scenario, since a fully
separable state admits perfect asymptotic PQSM if and only if it admits
perfect PQSM in the single-shot scenario. 

Using the above result, we will now prove the desired statement by
contradiction. Assume that the state $\rho_{\mathrm{sep}}^{ABC}$
with the above properties can be merged with some single-shot PPT
assisted LOCC protocol $\Lambda_{\mathrm{PPT}}$ between Bob and Charlie.
It then immediately follows that this protocol must merge each of
the states $\sigma_{i}^{BC}$ individually. Moreover, by convexity,
this protocol also merges each convex combination of the form 
\begin{equation}
\tau^{BC}=\sum_{i=0}^{14}q_{i}\sigma_{i}^{BC}.\label{eq:tau}
\end{equation}
Recall that the set of states of the form (\ref{eq:tau}) has finite
size within all two-qubit states. By convexity, this implies that
the protocol $\Lambda_{\mathrm{PPT}}$ can be used for single-shot
merging of \emph{any} state shared by Bob and Charlie. In particular,
this means that $\Lambda_{\mathrm{PPT}}$ can merge both states $\ket{00}^{BC}$
and $\ket{+0}^{BC}$. The existence of such a protocol would thus
imply that the states $\ket{0}$ and $\ket{+}=(\ket{0}+\ket{1})/\sqrt{2}$
can be perfectly teleported with the aid of PPT states on the single-copy
level. This is however impossible \citep{Henderson2000}, which is
the desired contradiction. This completes the proof that the aforementioned
family of states does not admit perfect asymptotic PQSM.

\section{States with vanishing asymptotic fidelity}

Here we will show that all states satisfying the inequality 
\begin{equation}
D_{\mathrm{LOCC}}^{A:BC}(\rho)>E_{n}^{AB:C}(\rho)=0\label{eq:vanishing-2}
\end{equation}
have zero fidelity in the asymptotic limit: 
\begin{equation}
{\cal F}_{\mathrm{PPT}}^{\infty}(\rho)=0.\label{eq:vanishing-3}
\end{equation}
For this we note that for all such states the final state $\sigma_{f}=\sigma_{f}^{ACR}$
is PPT with respect to the bipartition $A$:$CR$, and thus is nondistillable
with respect to this bipartition\footnote{Note that here the final state $\sigma_{f}^{ACR}$ is \emph{not} equal
to the initial state $\rho^{ABC}$ up to relabeling $B$ and $R$.}. This means that for any number of copies $n$ the fidelity of PQSM
is bounded above as follows: 
\begin{equation}
{\cal F}_{\mathrm{PPT}}(\rho^{\otimes n})=\sup_{\Lambda_{\mathrm{PPT}}}F(\sigma_{t}^{\otimes n},\sigma_{f})\leq\sup_{\tau\in\overline{{\cal D}}}F(\sigma_{t}^{\otimes n},\tau),\label{eq:F}
\end{equation}
where the final state $\sigma_{f}$ shared by Alice and Charlie is
given as $\sigma_{f}=\mathrm{Tr}_{B}[\Lambda_{\mathrm{PPT}}[\rho^{\otimes n}\otimes\rho^{R}]]$,
and the supremum in the last expression is taken over all states $\tau$
which are not distillable between Alice and Charlie. 

In the next step, we introduce the geometric distillability 
\begin{equation}
D_{g}(\nu)=1-\sup_{\tau\in\overline{{\cal D}}}F(\nu,\tau),
\end{equation}
and note that the target state $\sigma_{t}=\sigma_{t}^{ACR}$ in Eq.~(\ref{eq:F})
is distillable between Alice's system $A$ and Charlie's system $CR$.
For proving Eq.~(\ref{eq:vanishing-3}) it is thus enough to show
that for any distillable state $\nu$ the geometric distillability
approaches one in the asymptotic limit: 
\begin{equation}
\lim_{n\rightarrow\infty}D_{g}(\nu^{\otimes n})=1.\label{eq:geometric}
\end{equation}
Surprisingly, this is indeed the case for any distillable state $\nu$,
and the proof will be given in the following.

\section{Asymptotic geometric distillability}

In the following we consider the geometric distillability defined
as 
\begin{equation}
D_{g}(\rho)=1-\sup_{\sigma\in\overline{{\cal D}}}F(\rho,\sigma),
\end{equation}
where $F(\rho,\sigma)=\mathrm{Tr}(\sqrt{\rho}\sigma\sqrt{\rho})^{1/2}$
is the fidelity, and the supremum is taken over the set of nondistillable
states $\overline{{\cal D}}$. We will also consider the closely related
quantity 
\begin{equation}
D_{t}(\rho)=\inf_{\sigma\in\overline{{\cal D}}}T(\rho,\sigma),
\end{equation}
where $T(\rho,\sigma)=||\rho-\sigma||/2$ is the trace distance with
the trace norm $||M||=\mathrm{Tr}\sqrt{M^{\dagger}M}$. The trace
distance and fidelity are related as 
\begin{equation}
1-F(\rho,\sigma)\leq T(\rho,\sigma)\leq\sqrt{1-F(\rho,\sigma)^{2}}.\label{eq:FT}
\end{equation}

As we will show in the following, both quantities $D_{g}$ and $D_{t}$
are discrete in the asymptotic limit: asymptotically they attain only
the values 0 (if $\rho$ is nondistillable) and 1 (if $\rho$ is distillable).
For nondistillable states $\rho$ it is clear that $D_{g}$ and $D_{t}$
are both zero, and thus also zero asymptotically. We will now prove
the following equality for any distillable state $\rho$: 
\begin{equation}
\lim_{n\rightarrow\infty}D_{g}(\rho^{\otimes n})=\lim_{n\rightarrow\infty}D_{t}(\rho^{\otimes n})=1.\label{eq:asymptotic}
\end{equation}
Note that due to Eq.~(\ref{eq:FT}) it is enough to prove only one
of the equalities. In the following, we will prove the equality for
$D_{t}$.

In the first step, we note that Eq.~(\ref{eq:asymptotic}) is true
for the maximally entangled state $\ket{\phi^{+}}=(\ket{00}+\ket{11})/\sqrt{2}$.
This can be seen by noting that the fidelity between $\ket{\phi^{+}}^{\otimes n}$
and any nondistillable state $\sigma\in\overline{{\cal D}}$ is bounded
above as follows \citep{Horodecki1999a,Horodecki1999}: 
\begin{equation}
F(\ket{\phi^{+}}\bra{\phi^{+}}^{\otimes n},\sigma)\leq\frac{1}{2^{n/2}}.
\end{equation}
In the next step, note that for a distillable state $\rho$ there
exist a sequence of LOCC protocols $\Lambda_{n}$ acting on $n$ copies
of the state $\rho$ such that
\begin{equation}
\lim_{n\rightarrow\infty}T\left(\Lambda_{n}[\rho^{\otimes n}],\ket{\phi^{+}}\bra{\phi^{+}}^{\otimes\left\lfloor nE_{d}\right\rfloor }\right)=0,\label{eq:distillation-1}
\end{equation}
where $E_{d}$ is the distillable entanglement or $\rho$ and $\left\lfloor x\right\rfloor $
is the largest integer below $x$. Moreover, without loss of generality,
we assume that $\Lambda_{n}[\rho^{\otimes n}]$ and $\ket{\phi^{+}}^{\otimes\left\lfloor nE_{d}\right\rfloor }$
have the same dimension.

By applying the triangle inequality with some nondistillable state
$\sigma$ we further obtain:
\begin{align}
T\left(\ket{\phi^{+}}\bra{\phi^{+}}^{\otimes\left\lfloor nE_{d}\right\rfloor },\sigma\right) & \leq T\left(\Lambda_{n}[\rho^{\otimes n}],\ket{\phi^{+}}\bra{\phi^{+}}^{\otimes\left\lfloor nE_{d}\right\rfloor }\right)\nonumber \\
 & +T\left(\Lambda_{n}[\rho^{\otimes n}],\sigma\right).
\end{align}
Minimizing both sides of this inequality over all nondistillable states
$\sigma$, it follows that:
\begin{align}
D_{t}\left(\ket{\phi^{+}}^{\otimes\left\lfloor nE_{d}\right\rfloor }\right) & \leq T\left(\Lambda_{n}[\rho^{\otimes n}],\ket{\phi^{+}}\bra{\phi^{+}}^{\otimes\left\lfloor nE_{d}\right\rfloor }\right)\nonumber \\
 & +D_{t}\left(\Lambda_{n}[\rho^{\otimes n}]\right).
\end{align}

In the final step, we take the limit $n\rightarrow\infty$ and use
Eq.~(\ref{eq:distillation-1}), arriving at the following result:
\begin{equation}
\lim_{n\rightarrow\infty}D_{t}\left(\ket{\phi^{+}}^{\otimes n}\right)\leq\lim_{n\rightarrow\infty}D_{t}\left(\Lambda_{n}[\rho^{\otimes n}]\right).
\end{equation}
Recalling the fact that Eq.~(\ref{eq:asymptotic}) is true for the
maximally entangled state $\ket{\phi^{+}}$, this inequality implies
\begin{equation}
\lim_{n\rightarrow\infty}D_{t}\left(\Lambda_{n}[\rho^{\otimes n}]\right)\geq1.
\end{equation}
The proof of Eq.~(\ref{eq:asymptotic}) for all distillable states
is complete by noting that $D_{t}$ cannot increase under LOCC, i.e.,
$D_{t}(\rho^{\otimes n})\geq D_{t}(\Lambda_{n}[\rho^{\otimes n}])$.

\section{States with vanishing asymptotic fidelity have nonzero measure}

We will now show that the set of states with vanishing asymptotic
fidelity has nonzero measure in the set of all states. For this we
will present a family of three-qubit states $\rho=\rho^{ABC}$ which
are separable between $AB$ and $C$, do not touch the boundary of
separable states, and are distillable between $A$ and $BC$. This
assures that small perturbations of this state do not change its basic
properties, i.e., the perturbed states are also separable between
$AB$ and $C$, distillable between $A$ and $BC$, and thus have
vanishing asymptotic fidelity ${\cal F}_{\mathrm{PPT}}^{\infty}(\rho)=0$. 

The following three-qubit state has the aforementioned properties:
\begin{equation}
\rho=(1-p)\ket{\phi^{+}}\bra{\phi^{+}}\otimes\ket{0}\bra{0}+p\frac{\openone}{8}\label{eq:significant}
\end{equation}
with $\ket{\phi^{+}}=(\ket{00}+\ket{11})/\sqrt{2}$. The parameter
$p$ can be chosen from the range $0<p<p_{\max}$, and $p_{\max}>0$
is chosen such that the state $\rho$ is distillable between $A$
and $BC$ for all $p<p_{\max}$. 

In order to see that the state obtained in this way is not on the
boundary of separable states (with respect to the bipartition $AB$:$C$),
we consider a small perturbation of the form 
\begin{equation}
\rho'=\varepsilon\sigma+(1-\varepsilon)\rho
\end{equation}
 with an arbitrary three-qubit state $\sigma$. The proof is complete
by noting that for any $\sigma$ there exists some maximal parameter
$\varepsilon_{\max}(\sigma)>0$ such that $\rho'$ is separable for
all $0\leq\varepsilon\leq\varepsilon_{\max}(\sigma)$. This follows
directly from the existence of a separable ball around the maximally
mixed state~\citep{Gurvits2003}.

\bibliographystyle{apsrev4-1}
\bibliography{literature}

\end{document}